\documentclass[12pt]{article}
\begin{document}
\begin{center}
{\large { $\kappa$-MINKOWSKI SPACETIME THROUGH EXOTIC "OSCILLATOR"
}} \vskip 2cm
Subir Ghosh\\
\vskip .3cm
Physics and Applied Mathematics Unit,\\
Indian Statistical Institute,\\
203 B. T. Road, Calcutta 700108, India\\
\vskip .3cm
And\\
\vskip .3cm
Probir Pal\\
Physics Department,\\
Uluberia College,\\
Uluberia, Howrah 711315,  India.\\

\vskip 3cm
 {\bf{Abstract:}}
 We have proposed a generally covariant non-relativistic particle
 model that can represent the $\kappa$-Minkowski noncommutative
 spacetime. The idea is similar in spirit to the noncommutative particle
 coordinates in the lowest Landau level. Physically our model yields a novel
 type of dynamical system, (termed here as Exotic "Oscillator"), that obeys a
 Harmonic Oscillator like equation of motion with a {\it{frequency}} that is
 proportional to the square root of {\it{energy}}. On the other hand, the phase
 diagram does not reveal a closed structure since there is a singularity in the
 momentum even though energy remains finite. The generally covariant form is related to a generalization of the Snyder algebra in a specific gauge and yields the $\kappa $-Minkowski spacetime after a redefinition of the variables. Symmetry
 considerations are also briefly discussed in the Hamiltonian
 formulation. Regarding continuous symmetry, the angular momentum
 acts properly as the generator of rotation. Interestingly, both
 the discrete symmetries, parity and time reversal, remain intact
 in the $\kappa$-Minkowski spacetime.

\end{center}
\newpage
{\it Introduction}: Pointers from diverse areas in high energy
physics  indicate that one has to look beyond a {\it local}
quantum field theoretic description in the formulation of quantum
gravity. Very general considerations in black hole physics lead to
the notion of a fuzzy or Non-Commutative (NC) spacetime which can
avoid the paradoxes one faces in trying to localize a spacetime
point within the Planck length \cite{sz}. This is also
corroborated in the modified Heisenberg uncertainty principle that
is obtained in string scattering results. The recent excitement in
NC spacetime physics is generated from the seminal work of Seiberg
and Witten \cite{sw} who explicitely demonstrated  the emergence
of NC manifold in certain low energy limit of open strings moving
in the background of a two form gauge field. In this instance, the
NC spacetime is expressed by the Poisson bracket algebra (to be
interpreted as commutators in the quantum analogue),
\begin{equation}
\{x^\mu,x^\nu\}=\theta ^{\mu\nu}, \label{nc0} \end{equation} where
$\theta ^{\mu\nu}$ is a $c$-number constant. Up till now this form
of NC extension has been the popular one. However, notice that
Lorentz invarianc is manifestly violated in quantum field theories
built on this spacetime. Somehow, it appears that the very idea of
formulating field theories in this sort of spacetime,  consistent with quantum gravity, gets defeated by this pathology!

In a parallel development,  there have been intense activities in
studying other
 forms of NC spacetimes, such as the Lie algebraic form \cite{dfr}
 with  structure constants $C^{\mu\nu}_\lambda$,
\begin{equation}
\{x^\mu,x^\nu\}=C^{\mu\nu}_\lambda x^\lambda . \label{nc1}
\end{equation}
It is important to note that the  NC extension in (\ref{nc1}) is
{\it{operatorial}} \cite{dfr} and do not jeopardize the Lorentz
invariance in relativistic models, which is the case with
(\ref{nc0}) with constant $\theta ^{\mu\nu}$. (For an introduction
to this subject the readers are referred to \cite{ga}.) Of
particular importance in the above is a restricted class of
spacetimes known as $\kappa $-Minkowski spacetime (or $\kappa $-spacetime in short), that is
described by the algebra,
\begin{equation}
\{x_i,t\}=kx_i ~~,~~ \{x_i,x_j\}=\{t,t\}=0 .\label{kmin}
\end{equation}
In the above, $x_i$ and $t$ denote the space and time operators
respectively. Some of the important works in this topic that
discusses, among other things, construction of a quantum field
theory in $\kappa $-spacetime, are provided in \cite{g1,g2,g3}.
Very interestingly, Amelino-Camelia \cite{gam} has proposed an
alternative path to quantum gravity - "the doubly special
relativity" - in which {\it two} observer independent parameters,
(the velocity of light and Planck's constant), are present. It has
been shown \cite{kowal} that $\kappa$-spacetime is a realization
of the above. Furthermore, the mapping \cite{kowal} between
$\kappa$-spacetime and Snyder spacetime \cite{sny}, (the first
example of an NC spacetime), shows the inter-relation between
these models and "two-time physics" \cite{bars}, since the Snyder
spacetime can be derived from two-time spaces in a particular
gauge choice \cite{rom}. Our aim is to present a physically
motivated realization of the $\kappa$-spacetime.

An altogether different form \cite{sg} of NC phase space is
induced by spin degrees of freedom $S^{\mu\nu}$,
\begin{equation}
\{x^\mu,x^\nu\}=S^{\mu\nu}, \label{nc2}
\end{equation}
where once again the noncommutativity is operatorial and the model
is Lorentz invariant.

Now we come to the motivation of our work. In a non-relativistic
setup, NC space, originating from the lowest Landau level
projection of charged particles moving in a plane under the
influence of a uniform, perpendicular (and strong) magnetic field
\cite{sz}, has become the prototype of a simple physical system
(qualitatively) describing considerably more complex and abstract
phenomena, in this case open strings moving in the presence of a
background two form gauge field \cite{sw} mentioned before. Under
certain low energy limits, the mechanism by which NC manifolds
emerge in the string boundaries on the Branes, is similar to the
way NC particle coordinates appear in the Landau problem. This
sort of intuitive picture, if present, is very useful and
appealing. The NC space (or spacetime) one is talking about here
is of the form
\begin{equation}
\{x^\mu,x^\nu\}=\theta ^{\mu\nu}~~;~~\mid \theta \mid \sim \mid
B^{-1} \mid ~, \label{nc} \end{equation}
 where $\theta _{\mu\nu}$ is constant and the strength is
 proportional to the inverse of magnetic field $B$. Note that in
 the classical set up the commutators are interpreted as Poisson
 Brackets (or Dirac Brackets).

  In the phase space form of noncommutativity also
\cite{sg} there is a physical picture concerning spinning particle
models \cite{sg1} that induces the NC spacetime. However, such an
intuitive analogue is lacking for understanding the Lie algebraic
form of NC \cite{dfr}. Our present work is aimed at throwing some
light in this area.

In this paper we are going to put forward a non-relativistic
particle model that has an underlying phase space algebra
isomorphic to the $\kappa$-Minkowski one (\ref{kmin}). Hamiltonian
constraint analysis \cite{dir} reveals a novel dynamical system,
(termed here as  Exotic "Oscillator"): {\it{it has the square root
of energy as its frequency}}. This sort of feature is curiously
reminiscent of the quantum particle whose frequency is
proportional to its energy. Phase diagram analysis yields further
surprises, to be elaborated later.

However, demonstrating that the model truly represents the NC
$\kappa$-spacetime is not straightforward, the main
hurdle being the identification of the time operator
{\footnote{There is a version of $\kappa$-spacetime \cite{g5}
which has  only NC space coordinates. We intend to study this
model more closely in our framework since the NC time
complications will be absent here. We wish to thank the Referee
for pointing this out.}}. This requires a generalization of our
model to a generally covariant one \cite{ht}. The gauge invariance
(due to the symmetry under reparametrization of the evolution
parameter) allows us to choose a gauge condition that fixes the
time operator according to our requirement. This way of exploiting
a non-standard gauge condition to induce NC coordinates has been
used in \cite{stern} in constant spacetime noncommutativity
(\ref{nc}). \vskip .5cm \noindent {\it Mechanical model for
$\kappa$-spacetime}: We start by considering a canonical phase
space with the non-zero Poisson Brackets,
\begin{equation}
\{X_i,P_j\}=\delta _{ij}~~,~~\{\eta,\pi\}=1 \label{0}.
\end{equation}
The sets $(X_i,P_j)$ and $(\eta,\pi)$ are decoupled. (We do not
distinguish between upper and lower indices in the
non-relativistic setup.) Let us posit the following set of Second
Class Constraints (SCC) \cite{dir}
\begin{equation}
\chi_1\equiv \pi ~~~,~~~\chi_2\equiv \eta -k(\vec{P}.\vec{X})~.
 \label{1}
\end{equation}
 SCCs require
the usage of Dirac Brackets (DB) defined by,
\begin{equation}
\{A,B\}_{DB}=\{A,B\}-\{A,\chi _i\}\{\chi _i,\chi _j\}^{-1}\{\chi
_j,B\} \label{db}~,
\end{equation}
such that DB between an SCC and {\it{any}} operator vanishes. In
the present case, the simple constraint Poisson Bracket matrix and
its inverse are respectively,
\begin{equation}
\{ \chi_i,\chi_j\}=
 \left (
\begin{array}{cc}
 0 &  -1\\
1 &  0
\end{array}
\right ) \label{mat}
\end{equation}

\begin{equation}
\{ \chi_i,\chi_j\}^{-1}=
 \left (
\begin{array}{cc}
 0 &  1\\
-1 &  0
\end{array}
\right ) \label{-mat}
\end{equation}

 The non-vanishing  DBs are derived below:
\begin{equation}
\{X_i,\eta \}=kX_i ~~,~~\{P_i,\eta \}=-kP_i~~,~~\{X_i,P_j\}=\delta
_{ij}
 \label{2}
\end{equation}
Since we will always deal with DBs the subscript DB is dropped.

We now construct the following  Lagrangian that has the same SCC
structure as in(\ref{1}):
\begin{equation}
L= \frac{m}{2}  \vec{\dot X}^2 - 2cmk\eta (\vec{X}. \vec{\dot X})
+c\eta ^2+2mc^2k^2\eta ^2\vec{X}^2 .\label{3}
\end{equation}
$m$ denotes the mass of the non-relativistic particle and $c$ and
$k$ are two other constant parameters.

Reexpressed in the form,
$$L= \frac{m}{2}  \vec{\dot X}^2 +(2mc^2k^2\eta ^2+mck\dot \eta)\vec
{X}^2 + c\eta ^2 ,$$ one can think of the model as that of a
generalized form of oscillator whose spring coupling is not
constant and depends on $X_i$ itself through $\eta $. In fact
classically $\eta $ can be eliminated by solving the Gaussian to
yield a complicated non-linear model. Instead we prefer to work
with this polynomial form with the extra variable $\eta $. As we
shall see later, the model  describes a novel dynamical system.

 The conjugate momenta in (\ref{3}) are defined by,
 \begin{equation}
 P_i=m\dot X_i-c\eta ^2-2cmk\eta X_i ~~,~~\pi =0.
\label{mom}
\end{equation}
 The Primary constraint is
\begin{equation}
\chi_1\equiv \pi \approx 0 .\label{c1}
\end{equation}
Time persistence of $\chi_1$  generates the Secondary constraint
\begin{equation}
\chi_2\equiv \dot \chi_1=\{\chi_1,H\} \rightarrow \chi_2\equiv\eta
-k(\vec{P}.\vec{X})\approx 0 . \label{c2}
\end{equation}
These are the same as the constraints we started with at the
beginning in (\ref{1}). Obviously identical DBs as in (\ref{2})
will be reproduced. The Hamiltonian
\begin{equation}
H=\frac{\vec{P}^2}{2m}+2ck\eta (\vec{P}.\vec{X}) -c\eta ^2
 \label{4}
\end{equation}
 in the reduced phase space simplifies to the Exotic "Oscillator",
\begin{equation}
H=\frac{\vec{P}^2}{2m}+ck^2(\vec{P}.\vec{X})^2~.
 \label{5}
\end{equation}
It is worthwhile to emphasize the fact the model proposed here for
simulating $\kappa$-spacetime  has considerably more structure (in
the form of additional variables $\eta$ and $\pi$) than the
analogous model for constant noncommutativity \cite{sz}. This is
expected on the grounds that the Lie algebraic form of NC algebra
is non-linear and operatorial in nature. We also encounter
\cite{pro} similar complexities in analyzing a Lie algebraic
space-space NC algebra. \vskip .5cm \noindent {\it The Exotic
"Oscillator"}: The relevant Hamiltonian equations of motion are
\begin{equation}
\dot X_i=\{X_i,H\}=\frac{P_i}{m} +2ck^2(\vec{P}.\vec{X})X_i
~~,~~\dot P_i=-2ck^2(\vec{P}.\vec{X})P_i .\label{6}
\end{equation}
A further iteration in time derivative generates the following
Exotic "Oscillator" dynamics:
\begin{equation}
\ddot X_i=-w^2X_i \label{7}
\end{equation}
with $c=-b^2$ and the frequency $w$ identified as
\begin{equation}
w=\pm 2bk\sqrt H.\label{ho}
\end{equation}
Note the novel characteristic of dispersion where the frequency is
a function of the Hamiltonian or energy. This is curiously
reminiscent of the quantum mechanical dispersion $w\sim energy$.
This is one of the interesting results of the present analysis
{\footnote{ One might be tempted to think that tuning the exponent
in the $(\vec{P}.\vec{X})$ term in (\ref{5}), the quantum particle
dispersion can be obtained. However this is not the case as we
show in the Appendix A. As a curiosity, the quantum dispersion {\it will} be obtained if the Hamiltonian is proportional to $(H)^{\frac{3}{2}}$ with $H$ of the form of (\ref{5}). }}

From the above analysis, the Exotic "Oscillator" interpretation
seems to be straightforward, since for a particular value of the
energy, (which is a  conserved quantity), the "Oscillator" will
have a definite frequency given by (\ref{ho}). However, a phase
diagram of our model (see Figures 1 and 2) will reveal that the
above conclusion is {\it{not}} fully correct.

In the figures we have considered a simplified version of the
Hamiltonian (\ref{5}), in the "Oscillator" regime, in one space
dimension and all the parameters are taken to be unity,
\begin{equation}
H\equiv E=P^2(1-X^2). \label{pd}
\end{equation}
The phase diagram for $E=1$ is drawn in Figure-1. In Figure-2
phase diagrams are drawn for three values of energy $E=1,3,5$ and
they are compared with the Harmonic Oscillator Hamiltonian, for
the same set of energy values $E_{ho}=1,3,5$,
\begin{equation}
H\equiv E_{ho}=p^2+q^2. \label{pd1}
\end{equation}
In(\ref{pd}) and (\ref{pd1}), we use the parametric
representations respectively,
\begin{equation}
X=cos(r),~P=\frac{\sqrt{E}}{sin(r)}, \label{p}
\end{equation}
\begin{equation}
X=\sqrt{E}cos(r),~P=\sqrt{E}sin(r). \label{p1}
\end{equation}
It is evident that in (\ref{p}) there is a singularity at $r=0$.
Actually in the figures   we have plotted
$$X=cos(r),~P=\frac{\sqrt{E}}{a+sin(r)},$$ with $a=0.5$ in Figure-1 and $a=1.5$ in Figure-2.
The asymmetries in the figures are due to the choice of the value of $a$.
Indeed, in this model the momentum can diverge even though the
energy remains finite. It will be useful to construct a variant of
our model with the $a$-damping in-built. Clearly one has to be
more careful in interpreting $w$ containing the Hamiltonian
explicitly as frequency. It is clear from the Lagrangian in
(\ref{3}) that the model is qualitatively different from a
Harmonic Oscillator. An intuitive physical understanding of this
behavior of our Exotic "Oscillator", with the apparently simple
looking dynamics as depicted in (\ref{7}), is possible in the
Lagrangian version..

A Lagrangian framework is better suited to get the physical
picture corresponding to the Exotic "Oscillator". For the
one-dimensional model (\ref{pd}), exploiting the first order
formalism, we get the Lagrangian as,
\begin{equation}
 L\equiv P\dot X -H \label{k1}=
\frac{1}{2}\frac{\dot X^2} {(1-cX^2)} \label{k2}
\end{equation}
where we have eliminated $P$ using the equation of motion. The
impression is that of a "free" particle with an {\it{effective }}
mass. The singularity of this effective mass leads to the momentum
blowup{\footnote{The ad hoc momentum cut off $a$, introduced to
get a closed phase space diagram is in-built in the following
Lagrangian,
\begin{equation}
L=\frac{1}{2}\frac{(\dot X_i)^2}{(a+\sqrt{1-cX_i^2})^2}
.\label{k4}
\end{equation}.}}

The higher dimensional action is more involved:
\begin{equation}
L=P_i\dot X_i-H= \frac{1}{2}[(\dot X_i)^2+c\frac{(X_i\dot
X_i)^2}{1-cX_i^2}]\label{k3}.
\end{equation}
For a single space dimension, (\ref{k3}) reduces to
(\ref{k2}).\begin{equation} L=\frac{1}{2}\frac{(\dot
X_i)^2}{(a+\sqrt{1-cX_i^2})^2} \label{k41}
\end{equation}
Expressions similar to (\ref{k3}) have appeared in \cite{rom}.  It
will be very interesting if these models are related to known
physical systems. We will comment on this possibility at the end.

\vskip .5cm \noindent {\it Generally covariant framework}: Let us
now come to the main topic: $\kappa$-spacetime. From the
DBs (\ref{2}) it is evident that our aim is to identify the degree
of freedom $\eta $ as time. So far in this formulation, $\eta (t)$
is a (configuration space) degree of freedom just as $X_i(t)$ and
their evolution is dictated by the Hamiltonian in the conventional
way. Hence further work is to be done if $\eta $ is to identified
as time. Quite obviously, in $\kappa$-spacetime time is
an operator since it has non-trivial commutation relations. In our
classical scenario this will be reflected in the non-zero Dirac
Brackets concerning $\eta $.

Hence in order to identify $\eta $ as the time operator, the
natural way to proceed is to generalize the model to a
{\it{generally covariant}}  one \cite{ht}, which has more freedom
since the evolution is dependent on another parameter $\tau$ and
"time" is still not fixed or identified. In this formulation one
works in an extended phase space with one extra canonical pair
$\{X_0(\tau),P_0(\tau)\}=1$ and all the dynamical variables are
functions of the parameter $\tau$. The system is elevated to a
local gauge theory where the gauge symmetry is invariance under
reparametrizations of $\tau$. Therefore one has the freedom of
choosing a gauge condition in order to lift the above invariance
and this choice in effect can fix the time operator.
Conventionally $X_0(\tau)$ plays the role of the time operator and
normal Hamiltonian mechanics is recovered in the gauge
$X_0(\tau)=\tau$. In the present context our aim is to fix the
gauge so that the variable $\eta$ (introduced above) becomes the
time operator.

We follow \cite{ht} and reexpress the action $S$ of our model in
the following generally covariant form,
\begin{equation}
S=\int d\tau [P_i\dot X_i+\pi \dot\eta +P_0\dot X_0-H_E
].\label{10}
\end{equation}
The extended Hamiltonian,
\begin{equation}
H_E=u_0\phi _0+u_1\chi_1+u_2\chi _2 ,\label{11}
\end{equation}
has become a linear combination of constrains only and weakly
vanishes. In (\ref{11}) $\phi _0$ represents the FCC inducing
$\tau$-diffeomorphism,
\begin{equation}
\phi _0\equiv P_0+H \approx 0 ,
 \label{fcc}
\end{equation}
and $\chi_1$ and $\chi _2$ are the SCC-pair introduced at the
beginning in (\ref{1}) and $u_i$s denote multipliers. Note that in
this form we have reverted back to completely canonical
$(X_i,P_j)$ and $(\eta,\pi )$ phase space. However, for
convenience, we will exploit the partially reduced phase space
where the SCCs $\chi_1$ and $\chi_2$ are strongly zero with the
phase space algebra given in (\ref{2}). As mentioned before,
conventional dynamics, as obtained in (\ref{5},\ref{6}) is
recovered in the gauge $\phi _1\equiv X_0(\tau)-\tau\approx 0$
which constitutes the SCC pair together with $\phi _0$.

Now comes the most important part of our work. Since we are
interested in a particle model that generates the
$\kappa$-spacetime, we instead choose a gauge,
\begin{equation}
\phi _1\equiv X_0 -k(\vec{P}.\vec{X}) .
 \label{g}
\end{equation}
The reason for this choice is the following. Remember that we are
working in a truncated phase space where the SCC $\chi _2\equiv
\eta -kX_iP_i$ strongly vanishes and thus $\eta$ is already
identified with $kX_iP_i$. So in the above gauge (\ref{g}) the
time variable $X_0$ becomes related to $\eta $.

To get the fully reduced phase space, we now compute the secondary
set of DBs induced by the SCC pair $ (\phi _0,\phi _1)$ with
\begin{equation}
\{\phi_0,\phi_1\}=-(1-\frac{k\vec{P}^2}{m})\equiv -\alpha
.\label{13}
\end{equation}
We must remember to use the first set of DBs in (\ref{2}) as the
existing Bracket structure in the definition of DB in (\ref{db})
in the analysis at hand. This leads to the following more involved
final DB structure involving coordinate and momenta:
\begin{equation}
\{X_i,X_j\}=\frac{k}{m\alpha}(X_iP_j-X_jP_i)~,~~
\{X_i,P_j\}=\delta _{ij}+\frac
{k}{m\alpha}P_iP_j~,~~\{P_i,P_j\}=0~. \label{14}
\end{equation}
The algebra (\ref{14}) is a slightly more general form of the one
proposed by Snyder \cite{sny} due to the scaling by $\alpha $ in
the right hand sight. The pure form of Snyder algebra \cite{sny}
have appeared in \cite{rom} in a gauge fixed reduced two-time
model. The algebra with $X_0$ turns out to be,
\begin{equation}
\{X_i,X_0\}=\frac{kX_i}{\alpha}~~,~~ ~~\{P_i,X_0\}=-\frac
{kP_i}{\alpha}~.\label{14a}
\end{equation}
Notice that the spacetime, as obtained in (\ref{14},\ref{14a}), is
not the $\kappa$-spacetime that we set out to generate.
But this is rectified by introducing the following set of
variables
\begin{equation}
x_i\equiv
X_i-\frac{k}{m\alpha}(\vec{P}.\vec{X})P_i~~~,~~~p_i\equiv P_i
\label{15}
\end{equation}
in terms of which we obtain the following DBs
\begin{equation}
\{x_i,p_j\}=\delta _{ij}~~,~~\{x_i,x_j\}=0. \label{16}
\end{equation}
Hence, we will obtain identical dynamics as in (\ref{7},\ref{ho})
if in the general covariant framework we take the Hamiltonian,
\begin{equation}
H= H=\frac{p^2}{2m}+ck^2(\vec{p}.\vec{x})^2~. \label{17}
\end{equation}
This is obtained from (\ref{5}) by replacing the set $(X_i,P_j)$
by $(x_i,p_j)$ {\footnote{Another derivation of the Hamiltonian
operator is provided in Appendix B where the second stage DBs
(induced by the pair $(\phi _0,\phi _1)$) are not needed.}}.

To get a representation of the time operator, we note that,
\begin{equation}
\{x_i,X_0\}=\{X_i-\frac{k}{m}(\vec{P}.\vec{X})P_i,X_0\}=\frac{k}{\alpha}(X_i+\frac{k}{m}(\vec{P}.\vec{X})P_i).
\label{18}
\end{equation}
However, the correct DB for $\kappa$-spacetime is
generated with the time variable
\begin{equation}
t\equiv k\alpha (\vec{P}.\vec{X}), \label{18a}
\end{equation}
for which we obtain,
 \begin{equation}
\{x_i,t\}=kx_i. \label{19}
\end{equation}
The operator conjugate to the time is obtained below:
\begin{equation}
\{t,\frac{1}{2\kappa}ln \vec P ^2\}=1. \label{cont}
\end{equation}
This constitutes our final result. We also note that the $k=0$
limit that reduces $\kappa$-Minkowski to commutative spacetime is
smooth everywhere. \vskip .5cm \noindent {\it Exotic "Oscillator"
in Snyder space}: Because of the non-linearity involved in the
Snyder algebra (\ref{14}), probably one of the simplest but
interesting mechanical model in Snyder space is the Exotic
"Oscillator". Consider the Snyder algebra,
\begin{equation}
\{X_i,X_j\}=-\gamma(X_iP_j-X_jP_i)~,~~
\{X_i,P_j\}=\delta _{ij}-\gamma P_iP_j~,~~\{P_i,P_j\}=0~, \label{14b}
\end{equation}
(where for later convenience we have taken the NC $\kappa$-parameter to be $-\gamma $).
For small $\gamma$ one finds the following set of equations of motion,
\begin{equation}
\ddot X_i=-2\gamma HX_i~~,~~~\ddot P_i=-2\gamma HP_i~~,
 \label{ex1}
\end{equation}
for the Hamiltonian,
\begin{equation}
H=\frac{1}{2}X_iX_i+\frac{\gamma}{2}(X_iP_i)^{2}.
 \label{ex2}
\end{equation}
The above equations are valid up to $O(\gamma )$. The choice of the sign of $\gamma$ is tuned
to get the dynamics in the Exotic Oscillator form. Notice the difference between the
Hamiltonian (\ref{ex2}) and the Hamiltonian (\ref{15}) where $\{X_i,P_j\}=\delta_{ij}$
as in (\ref{2}). Thus effectively the set $\{X_i,P_j\}$ is the canonical set $\{x_i,p_j\}$
in our notation. To $O(\gamma )$, one can recover the Hamiltonian (\ref{17}) from (\ref{ex2})
by exploiting the mapping (\ref{s2}) given below.
\vskip .5cm \noindent {\it Symmetries}:
Symmetry principles are playing increasingly major roles in
contemporary physics. The fate of conventional spacetime
symmetries in the context of NC theories is an important issue
since one is changing the underlying spacetime structure itself.
Poincare invariance in the canonically NC field theories is
explicitly broken \cite{sub}. However, in $\kappa$-spacetime,
there appears a {\it deformation} of Lorentz symmetry
\cite{g6,kowal}. These issues are more pertinent where
relativistic field theories are concerned. Hence we will restrict
ourselves to the symmetries that are relevant for non-relativistic
(Hamiltonian) quantum mechanics. We will find non-trivial
differences from the results obtained in  \cite{bal} where a
space-time constant ({\it i.e.} canonical) form of
noncommutativity has been considered.

We start with the angular momentum operator $L_{ij}=X_iP_j-X_jP_i$
and find
\begin{equation}
\{L_{ij},X_k\}=\delta_{ik}X_j-\delta_{jk}X_i~;~~\{L_{ij},P_k\}=\delta_{ik}P_j-\delta_{jk}P_i
. \label{s1}
\end{equation}
With $l_{ij}=x_ip_j-x_jp_i $ this is isomorphic to the conventional algebra,
\begin{equation}
\{l_{ij},x_k\}=\delta_{ik}x_j-\delta_{jk}x_i~;~~\{l_{ij},p_k\}=\delta_{ik}p_j-\delta_{jk}p_i
. \label{s11}
\end{equation}
It should be remembered that  $(X_i,P_j)$ obey the generalized
algebra (\ref{14}) whereas $(x_i,p_j)$ obey the commutative
spacetime algebra (\ref{16}). This shows that Snyder phase space
behaves canonically under rotations. In fact, exploiting the
inverse mapping of (\ref{15}),
\begin{equation}
X_i\equiv
x_i+\frac{k}{m\alpha}(\vec{p}.\vec{x})p_i~~~,~~~P_i\equiv p_i ,
\label{s2}
\end{equation}
it is easy to see that
\begin{equation}
L_{ij}=X_iP_j-X_jP_i=x_ip_j-x_jp_i\equiv l_{ij}. \label{s3}
\end{equation}
Interestingly, we find the {\it {time operator to be rotationally
invariant}},
\begin{equation}
\{L_{ij},X_0\}=0~,~~~\{l_{ij},t\}=0. \label{s4}
\end{equation}
Hence, unlike the case discussed in \cite{bal}, $L_{ij}$ can
regarded as the generator of spatial rotations in $\kappa
$-spacetime.

Let us now turn to the discrete symmetries of the quantum theory
where we identify $\{,\}\Rightarrow -i[,]$ and
$p_i=-i\frac{\partial}{\partial x_i}$. Considering parity
transformations in $\kappa$-spacetime,
\begin{equation}
P:~~t\rightarrow t~,~~x_i\rightarrow -x_i~, \label{s5}
\end{equation}
we find the NC commutation relations
\begin{equation}
[x_i,x_j]=0~,~~[x_i,t]=i\kappa x_i ~,~~i\rightarrow i~, \label{s6}
\end{equation}
are preserved under parity, where $P$ is a linear operator. At the
same time, considering the time reversal operator $T$ as an
anti-linear operator, we find that the transformations
\begin{equation}
T:~~t\rightarrow -t~,~~x_i\rightarrow x_i~,~~i\rightarrow -i~,
\label{s7}
\end{equation}
 preserve (\ref{s6}) as well. {\it {Hence $P$ and $T$ symmetries remain intact in
 $\kappa$-spacetime}}. For charge conjugation invariant models based
 on $\kappa $-spacetime algebra, $CPT$ will remain a valid
 symmetry. Once again we note
 the crucial difference
with canonical NC spacetime results \cite{bal}. \vskip .5cm
\noindent
{\it {Conclusion and future outlook}}:\\
\noindent We have succeeded in presenting a non-relativistic
particle model which reproduces the $\kappa$-spacetime.
The spirit of our work is in analogy with  (the lowest level
projection of ) Landau problem of charges moving in  a plane in a
perpendicular magnetic where the particle positions become
effectively noncommutative  with constant $\theta$. In the
process, we have found that physically our particle model yields a
novel type of dynamics that appears to be "Oscillator"-like with a
frequency proportional to the square root of the energy.
Surprisingly the motion is not truly periodic which is revealed in
the study of the phase diagram. Subsequently a generalization of
the model to a generally covariant one leads to a definition of
time that gives the full $\kappa$-Minkowski algebra. Furthermore,
we have shown how a generally covariant reformulation of the model
describes the (generalized) Snyder spacetime in a particular gauge
and eventually leads to the $\kappa $-Minkowski spacetime. Study
of the continuous (rotational) and discrete (parity and time
reversal) symmetries reveal that the $\kappa $-Minkowski spacetime
is probably a better option than the constant space-time
noncommutativity, as studied in \cite{bal}. This is primarily
because angular momentum is the correct generator for rotations
and parity and time reversal symmetries are kept intact. Hence
maintaining $CPT$-invariance will not pose any problem.

We note some points that are to be studied in future. In the
Exotic "Oscillator" context, a physical interpretation of the open
phase diagram is required. One can try to construct an extension
of the model, with the characteristic features as we have noted,
but having at the same time  a closed periodic motion. It will be
very interesting to quantize the model. Also it would be
interesting to investigate the type of systems that can induce
quantum particle like dispersion and to study the kind of
spacetimes they represent. Similar analysis, as has been performed
here, for the general Lie algebraic noncommutative spacetime is
under study.

In the context of obtaining the $\kappa$-Minkowski spacetime from
our model, one can exploit an alternative framework (see Banerjee
et.al. in \cite{stern}) where the identification of the time
operator might be more direct. There is the possibility that some
familiar interacting model, in a {\it{non-standard gauge}}, will
be equivalent to the  particle model proposed here. As a more
ambitious programme, taking a cue from the Landau problem - string
analogy in the context of noncommutative spacetime, one can try to
construct string models yielding $\kappa$-Minkowski spacetimes.
Our Exotic "Oscillator" model can help in the construction of the
latter. A positive indication in this direction is that in a
relativistic generalization  of the present model the de Sitter
metric plays a pivotal role and it is indeed natural to extend the
framework for strings moving in de Sitter background. These
results will be reported elsewhere \cite{subg}.

\vskip .5cm \noindent
{\bf{Appendix A:}}\\
For a more general form of the Hamiltonian, comprising of
canonical $(x_i,p_j)$ variables,
\begin{equation}
\tilde H=\frac{(\vec{p})^2}{2m}+c(\vec{p}.\vec{x})^n \label{20}
\end{equation}
we obtain the following equation of motion,
\begin{equation}
\ddot
x_i=nc(\vec{p}.\vec{x})^{n-2}[(n-1)\frac{(\vec{p})^2}{m}+nc(\vec{p}.\vec{x})^n]x_i
. \label{21}
\end{equation}
It is easy to see that only $n=2$ reproduces the exotic
oscillator. \\
{\bf{Appendix B:}}\\From the weakly vanishing Hamiltonian,
\begin{equation}
H=u\phi _0 \label{22}
\end{equation}
and the explicitly time ($\tau $) dependent gauge condition $\phi
_1\equiv X_0 -k(\vec{P}.\vec{X}) -\tau$, time persistence of $\phi
_1$ determines the multiplier $u$ in (\ref{22}) in the following
way:
\begin{equation}
\frac{d\phi _1}{d\tau}=\frac{\partial\phi_1}{\partial\tau}+\{\phi
_1,\phi _0\}~\rightarrow ~u=\frac{1}{\alpha}. \label{23}
\end{equation}
The equations of motion (modulo constraint) are
\begin{equation}
\dot X_i=\{X_i,\frac{\phi
_0}{\alpha}\}=\frac{1}{\alpha}(\frac{P_i}{m}
+2ck^2(\vec{P}.\vec{X})X_i) ~~,~~\dot
P_i=-\frac{2ck^2}{\alpha}(\vec{P}.\vec{X})P_i . \label{24}
\end{equation}
For low velocity (or large mass) $\alpha \approx 1$ the dynamical
equations of (\ref{6}) are reproduced.

\vskip .5cm \noindent
 {\bf{Acknowledgements:}} It is a
pleasure to thank Dr. Rabin Banerjee for discussions. S.G. is
grateful to Dr. Giovanni Amelino-Camelia for initiating his
interest in $\kappa$-Minkowski spacetime. I am grateful to the
Referee for the constructive comments.

\end{document}